\documentclass[aps,pra,twocolumn,showpacs,groupedaddress,10pt]{revtex4-1}
\usepackage{graphicx}
\usepackage{amsmath}
\usepackage{amssymb}
\usepackage{verbatim}
\usepackage{rotating}

\begin{document}

\title{Self-Correcting Quantum Memory in a Thermal Environment}

\author{Stefano Chesi, Beat R\"othlisberger, and Daniel Loss}
\affiliation{Department of Physics, University of Basel, Klingelbergstrasse 82, CH-4056 Basel, Switzerland}
\date{\today}

\begin{abstract}
The ability to store information is of fundamental importance to any computer, be it classical or quantum. To identify systems for quantum memories which rely, analogously to classical memories, on passive error protection (`self-correction') is of greatest interest in quantum information science. While systems with topological ground states have been considered to be promising candidates, a large class of them was recently proven unstable against thermal fluctuations. Here, we propose two-dimensional (2D) spin models unaffected by this result. Specifically, we introduce repulsive long-range interactions in the toric code and establish a memory lifetime polynomially increasing with the system size. This remarkable stability is shown to originate directly from the repulsive long-range nature of the interactions.
We study the time dynamics of the quantum memory in terms of diffusing anyons and support our analytical results with extensive numerical simulations. Our findings demonstrate that self-correcting quantum memories can exist in 2D at finite temperatures.
\end{abstract}


\maketitle

\section{Introduction}

Quantum computers 
cannot be realized without the help of error correction \cite{Nielsen2000}. 
By encoding quantum information into logical states and designing correction circuits
working on them,
computations and information can in principle be protected from decoherence. However, the need for such an active control mechanism 
poses a major challenge for any physical implementation.
It is therefore of greatest interest to look for passively protected systems which are intrinsically stable against the destructive influence of a thermal environment.
For this reason, the idea to encode quantum information in a topologically
ordered ground state $|\Psi_0\rangle$ of a suitable
Hamiltonian has attracted a lot of interest \cite{Kitaev2003, Dennis2002, Kitaev2006, Bacon2006,Trebst2007, Tupitsyn2008, Vidal2009a, Vidal2009b, Nussinov2008, Alicki2009, Iblisdir2009, Iblisdir2010,  Bravyi2009, Kay2008, Pastawski2009, Pastawski2009b, Hamma2009, Chesi2009a}.

Important candidates among such topological models are stabilizer Hamiltonians \cite{Nielsen2000,Gottesman1997},
which are given by a sum of mutually commuting many-body Pauli operators. The
advantage of such Hamiltonians is that the full energy spectrum is known
and error correction schemes are readily derived \cite{Nielsen2000,Gottesman1997}. 
A prototypical example of such models is the toric code proposed in Ref.~\cite{Kitaev2003}, 
for which the stability against Hamiltonian perturbations \cite{Trebst2007, Tupitsyn2008, Vidal2009a, Vidal2009b, Pastawski2009b} and thermal fluctuations \cite{Dennis2002, Nussinov2008, Alicki2009, Iblisdir2009, Iblisdir2010} was studied extensively.
However, recent results \cite{Bravyi2009, Kay2008} show that in one and
two spatial dimensions no stabilizer Hamiltonian with finite-range interactions (including the
toric code model) can serve as a self-correcting quantum memory due to the 
errors induced by a thermal environment.

In other words, increasing the size of such a system does not
prolong the protection of its ground-state space from decoherence. These negative
results point toward the fundamental question whether topologically ordered
quantum states, and hence self-correcting quantum memories, can exist at all
on a macroscopic scale. In the following, we will demonstrate that
self-correcting properties of two-dimensional (2D) stabilizer Hamiltonians can indeed be
established when we allow for long-range repulsive interactions between
the elementary excitations (anyons). While the purpose of the present work
is of principal nature, we note that such interacting models can be expected to be realized in physical systems.
We discuss this issue in greater detail at the end, where we also show how tunable repulsive long-range interactions could be mediated via photons in an optical cavity.

The outline of the paper is as follows: In Sec.~\ref{sec:model} we
introduce a toric code model with repulsive long-range interactions between anyons. 
In Sec.~\ref{app:simulation} we describe how to simulate the dynamics of the model in contact with Ohmic or super-Ohmic thermal baths. A discussion of the
decoherence caused by anyon diffusion and an expression of the
memory lifetime as a function of system parameters is provided in Sec.~\ref{sec:lifetime}. The main results of our paper are in Secs.~\ref{sec:mf} and \ref{sec:lifetime_interacting} where, first by an analytical mean-field treatment 
and then by direct numerical simulation, we demonstrate the self-correcting 
properties of our model. Section~\ref{sec:long_range} contains a discussion of the possible implementations of the long-range anyon interaction and Sec.~\ref{sec:conclusion} concludes the paper with our final remarks. 

\begin{figure}
 \includegraphics[width=0.95\columnwidth]{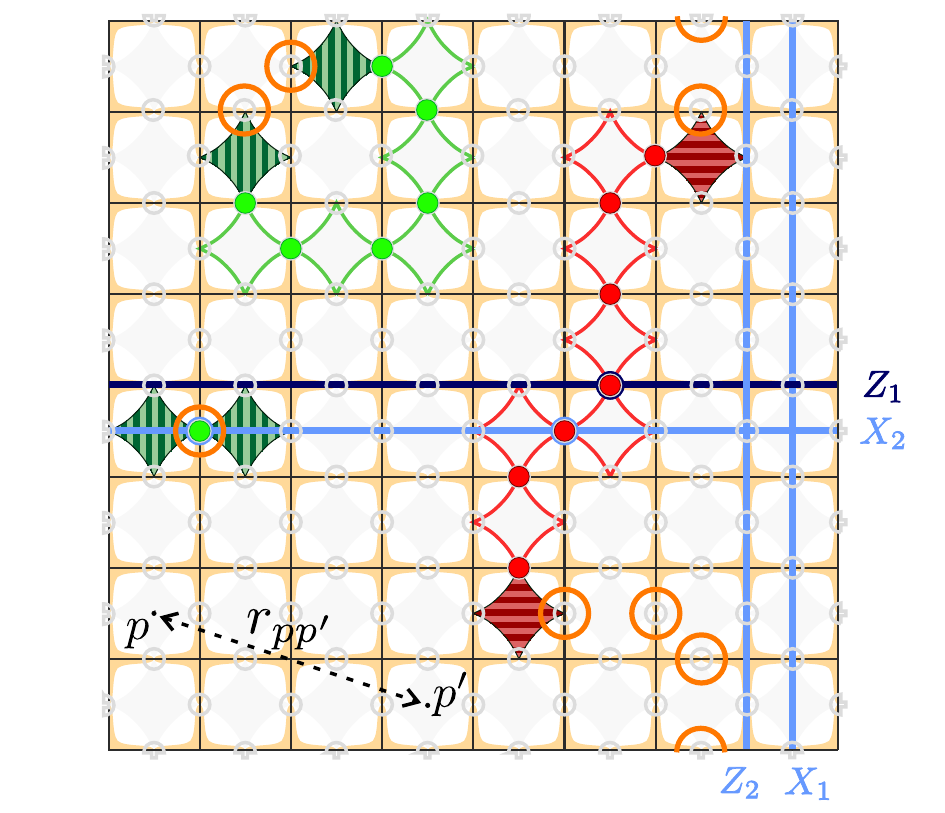}
 \caption{(Color online.) Quantum memory based on the toric code. Illustrated is an $8 \times 8$ lattice (periodic boundary conditions) with a total of 128 spins-$\frac{1}{2}$ [gray (smaller) circles] on its edges. The four-body plaquette and star operators are indicated in the background. A particular choice for all logical operators $X_1$, $Z_1$, $X_2$, and $Z_2$ is shown, although we will focus only on the decay of $Z_1 \equiv Z$ (see main text). A number of spins is affected by $\sigma_x$-errors (solid dots), leading to excited plaquettes, or `plaquette anyons' (striped plaquettes). Measuring the plaquette operators yields the positions of the excited plaquettes, but reveals no information about how they were originally paired or which path (indicated by the framed plaquettes) they took.
A minimum-weight error correction procedure (see Sec.~\ref{app:simulation_details}) applies $\sigma_x$-operators to the spins marked by the larger orange circles. While the vertically striped green anyons are annihilated `properly' (with a trivial loop of errors remaining from the top pair and no error from the bottom pair), the horizontally striped red pair is connected around a topologically non-trivial loop on the torus. Although this last pair is annihilated as well, an uncorrected $\sigma_x$-error remains on the logical $Z$ string, having thereby introduced a logical error in the state stored in the memory.} \label{fig:toric code}
\end{figure}

\section{Repulsive long range interactions in the Kitaev model}\label{sec:model}

The model under study is defined on a $L\times L$ square lattice with periodic boundary conditions (a `torus'), and a spin-$\frac{1}{2}$ is placed on each of its $2L^2$ edges. Starting from the toric code model \cite{Kitaev2003}, we consider the more general stabilizer Hamiltonian
\begin{equation}\label{H0}
H_0= \frac12 \sum_{p p'} U_{pp'} n_p n_{p'} + \frac12 \sum_{s s'} V_{ss'} n_s n_{s'},
\end{equation}
where $n_p=(1-\prod_{i\in p} \sigma_{z,i})/2$, $n_s=(1-\prod_{i\in s} \sigma_{x,i})/2$, and $\sigma_{x, i}, \sigma_{z, i}$ denote the usual single-spin $x$ and $z$ Pauli operators applied to spin $i$. The indices $p$ and $p'$ run over all `plaquettes' (involving the four spins on the edges of a unit cell), whereas $s$ and $s'$ run over all `stars' (involving the four spins around a corner of a unit cell), see Fig.~ \ref{fig:toric code}. The operator $n_p$ ($n_s$) has eigenvalues $0, 1$ and counts the number of plaquette- (star-) anyons at site $p$ ($s$). The fourfold degenerate energy levels encode two qubits with logical operators given by $Z_i = \prod_{k\in \ell_i} \sigma_{z,k}$ and $X_i = \prod_{k\in \ell'_i} \sigma_{x,k}$, $i = 1, 2$, where $\ell_i$ and $\ell'_i$ are strings of spins topologically equivalent to single loops around the torus (see Fig. \ref{fig:toric code} for an example). These operators commute with all $n_p$ and $n_s$ and obey themselves the usual spin commutation relations. 

Note that by specializing to $U_{p p'} = 2J\delta_{p p'}$ and $V_{s s'} = 2J\delta_{s s'}$, where $J > 0$ is the single-anyon excitation energy, the Kitaev original toric code model is recovered. Except for the boundary conditions, the structure of the toric code is very similar to an earlier model by Wegner \cite{Wegner1971, Kogut1979}. Wegner's Ising lattice gauge theory involves only plaquette operators in the Hamiltonian ($U_{p p'} = 2J\delta_{p p'}$ and $V_{s s'} = 0$), while the stars play the role of a gauge symmetry group. Both the Kitev Hamiltonian and the two-dimensional Wegner model have no finite-temperature phase transition, as can be obtained by mapping them to one-dimensional Ising chains \cite{Wegner1971,Kogut1979, Dennis2002, Nussinov2008}. Finally, the Kitaev model is also equivalent to a model proposed later by Wen \cite{Wen2003,Nussinov2007}.

Since all $n_p$ and $n_s$ are mutually commuting, the Hamiltonian Eq. \eqref{H0} describes two independent lattice gases of plaquettes and stars, respectively. Without loss of generality, we can thus restrict our analysis to the dynamics of plaquettes and their influence on one of the $Z_i$ operators, say $Z_1 \equiv Z$. A corresponding logical operator $Z_{\rm ec}$ is defined by the error correction procedure (see Fig.~\ref{fig:toric code} and Sec.~\ref{app:simulation_details}). Consequently, we set $V_{ss'}=0$ for all stars while assuming the plaquette interactions $U_{p p'}$ to be of the generic form
\begin{equation}\label{Upp_long_range}
U_{pp'}=2J \delta_{pp'}+\frac{A}{(r_{pp'})^\alpha} (1-\delta_{pp'}),
\end{equation}
where $r_{pp'}$ denotes the shortest distance on the torus between
the centers of plaquettes $p$ and $p'$, see Fig. \ref{fig:toric
code}. The strength of the repulsive plaquette interaction is given by the
energy $A \geq 0$, and the interaction is long-range for $0 \leq \alpha <
2$ (see below). The model is also equivalent to a long-range Ising model, see Appendix~\ref{app:ising_chain}. The case of a positive logarithmically diverging interaction (which results in attractive forces between the anyons \cite{Dennis2002}) was recently discussed in Ref.~\cite{Hamma2009}.

\section{Error models and simulations} \label{app:simulation}

\subsection{Error models}

We model the interaction of the system with a thermal environment by coupling each spin to a bath which can introduce $\sigma_x$-errors \cite{footnote01} in the initial state $|\Psi_0 \rangle$, assumed to be a ground state of Eq.~(\ref{H0}). From a standard master equation approach in the weak coupling limit \cite{Davies1974, Alicki2009}, we derive a rate equation for the probabilities $p_m$ of the system to be in state $|\Psi_m\rangle = \prod_{i\in m} \sigma_{x, i} |\Psi_0\rangle$, where $\{m\}$ is the set of all possible patterns of $\sigma_x$-errors. This rate equation reads
\begin{equation}\label{rate equation}
\dot p_m = \sum_{i}\left[\gamma(-\omega_i(m)) p_{x_i(m)} - \gamma(\omega_i(m)) p_m\right],
\end{equation}
where we have defined $x_i(m)$ to be the state $m$ with an additional $\sigma_x$-error applied to spin $i$, and $\omega_i(m) = \epsilon_m - \epsilon_{x_i(m)}$ is the energy difference between the states $m$ and $x_i(m)$.
The time evolution of the probabilities $p_{m}$ determines the decay of the expectation values $\langle Z_{(\rm ec)} \rangle = \sum_m p_m \langle \Psi_m |Z_{(\rm ec)}| \Psi_m \rangle$.

The rates $\gamma(\omega)$ describe the transition probabilities between states with energy difference $\omega$. A standard expression for $\gamma(\omega)$ can be obtained from a spin-boson model and reads \cite{Leggett1987, DiVincenzo2005}
\begin{equation}\label{gamma_bath_text}
\gamma(\omega)=2 \kappa_n \left|\frac{\omega^n}{1-e^{-\beta\omega}}\right| e^{-|\omega|/\omega_c}.
\end{equation}
Here, $\beta = 1/T$, with $T$ being the temperature of the bath (we
set Boltzmann's constant to one). For simplicity, we assume in the following a large cut-off energy $\omega_c \to \infty$. For $n = 1$, the bath is
called `Ohmic', whereas for $n \geq 2$ it is called `super-Ohmic'.
We find in this work that $n$ has a strong influence on the decay
times of the encoded states, with super-Ohmic baths providing the
best scaling of the memory lifetime with $L$. These are not uncommon and emerge, e.g., 
for quantum dot spins in contact with phonons~\cite{Golovach2004}. 

 \begin{figure}
\includegraphics[width = 0.95\columnwidth] {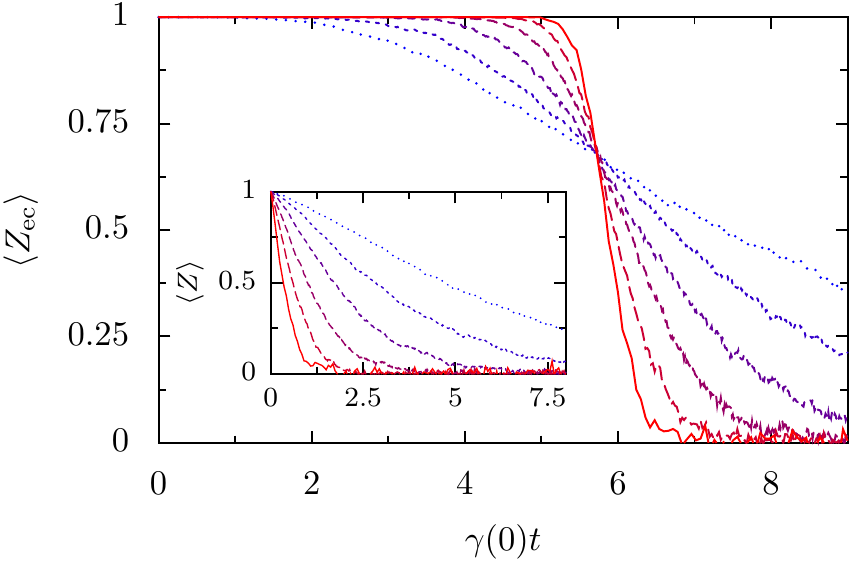}
 \caption{\label{nonint_fig} (Color online.) Decay of the logical $Z$ operator in the non-interacting toric code. The simulation data is obtained for grid sizes
$L$ increasing by powers of two from 16 (dotted blue) to 512 (solid red). All curves are ensemble averages over $10^4$ runs. The main plot displays $\langle Z_{\rm ec}\rangle$, which is the average value of $Z$ one would find if an error correction scheme would be applied at the readout time $t$. The inset shows the expectation value of the bare (uncorrected) logical $Z$ operator. We have used $T/J=0.3$, and $\gamma(0)=\gamma(2J)$. See Sec.~\ref{app:simulation_details} for further details on the simulation.}
 \end{figure}

\subsection{Simulations and error correction} \label{app:simulation_details}

The eigenstates of Eq.~(\ref{H0}) are highly entangled, but it is nevertheless possible to 
perform classical simulations of the quantum memory in the simple framework discussed above. In order to achieve a
time evolution in accordance with Eq.~(\ref{rate equation}), each
iteration of a simulation consists of the following steps. (i) We
record the relevant parameters of the system. (ii) We calculate the
total spin flip rate $R = \sum_i \gamma(\epsilon_s -
\epsilon_{x_i(s)})$, where $s$ is the current state of the system.
(iii) We draw the time $\Delta t$ it takes for the next spin to flip
from an exponential distribution, $\Delta t \sim {\rm Exp}(1/R)$,
and then add this to the current total time. (iv) We calculate all
individual spin flip probabilities $p_i = \gamma(\epsilon_s -
\epsilon_{x_i(s)})/R$ and flip a spin at random accordingly. After
some initially specified time has been reached, we stop and have
obtained a single `run'. The final data presented in this work is
then generated by averaging over many (typically several thousand) runs.

Although continuous monitoring and error-correction are not required in a passive memory during the storage time, it is still beneficial to apply an error correction scheme once the memory is being read out. By $\langle Z_{\rm ec}\rangle(t)$, we denote in this work the average value of $Z$ we would have obtained if we had performed error correction at time $t$. The goal here is to properly annihilate corresponding anyons (by applying $\sigma_x$-operations), thereby reverting the undesired operations performed by anyon paths crossing the logical operator strings. However, since only the positions of the anyons are known, this correspondence has to be guessed. We do this by choosing the pairing with the minimal sum of connection path lengths using \texttt{Blossom~V}~\cite{Kolmogorov2009}, which is the latest improvement on Edmonds' minimal-weight perfect matching algorithm \cite{Edmonds1965}. If many anyons are present, using the complete graph as the input to this algorithm is numerically infeasible. In excellent approximation, we therefore replace the complete graph by a Delaunay triangulation \cite{footnote03}.

As a useful reference, we show in Fig.~\ref{nonint_fig} numerical results for the non-interacting system, i.e., $A = 0$. The relevant rates entering Eq. \eqref{rate equation} are $\gamma(0)$ (rate for an anyon to hop to a free neighboring site), $\gamma(-2J)$ (rate to create an anyon pair) and $\gamma(2J) = \gamma(-2J)e^{2J\beta}$ (rate to annihilate a pair of adjacent anyons, obtained from the detailed balance condition). Figure~\ref{nonint_fig} illustrates the typical behavior of $\langle Z \rangle$ and $\langle Z_{\rm ec} \rangle$, in agreement with previous literature \cite{Alicki2009,Nussinov2008,Hamma2009, Bravyi2009,
Kay2008}. We refer to Sec.~\ref{sec:lifetime} for a more detailed discussion.

\section{Diffusion of anyons and memory lifetime}\label{sec:lifetime}

\begin{figure}
\includegraphics[width=0.95\columnwidth]{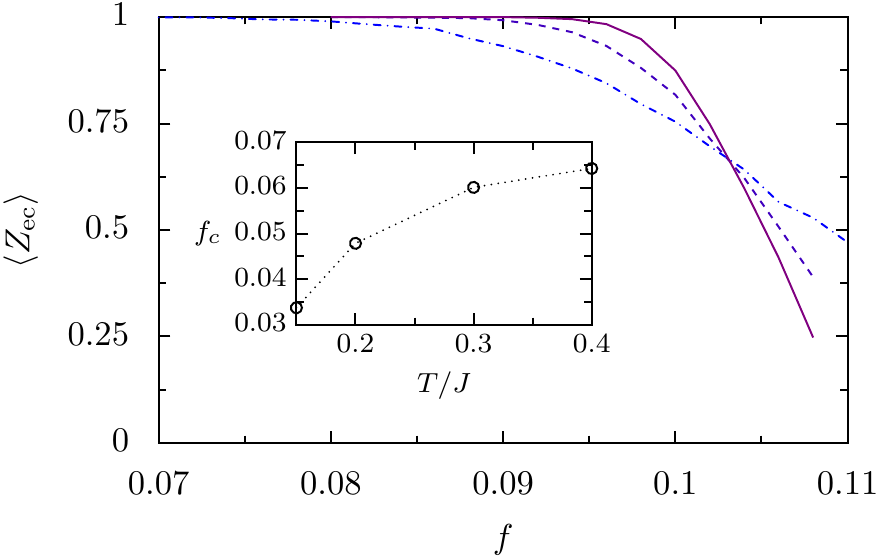}
    \caption{\label{Z_vs_f} (Color online.) Average of the corrected operator $Z_{\rm ec}$ for a model with independent $\sigma_x$-errors occurring with probability $f$ at each spin. The dashed-dotted, dashed, and solid curves refer to our numerical simulations with lattice sizes $L=40, 100, 200$, respectively. The error correction fails at a value $f_c\simeq 0.1$, which is slightly smaller than the value $0.11$ from Ref.~\cite{Dennis2002}. In the inset, we plot the value of $f_c$ from simulations of the non-interacting toric code in contact with a bath at temperature $T$  and $\gamma(0)=\gamma(2J)$. The fraction $f_c$ is extracted at the time $\tau$ when $\langle Z_{\rm ec} \rangle$ decays to zero in the limit of large $L$ (see Fig.~\ref{nonint_fig}). This value is always smaller than $f=0.11$ and depends on $T$.}
\end{figure}

It is the purpose of this section to establish a formula for the lifetime of the quantum memory. A static criterion was discussed in Ref.~\cite{Dennis2002}: assuming independent errors, the toric code can be mapped to a random-bond Ising model, and a threshold probability $f_c = 0.11$ is obtained. In the thermodynamic limit, retrieval of the encoded information is impossible if the relative number of errors is above this value. Below $f_c$, recovery is achieved with probability one. Numerically, we find a similar threshold $f_c\approx 0.1$ for the same error model, see Fig.~\ref{Z_vs_f}. This shows that our implementation of the minimum-weight error correction scheme works close to optimal.

Also in the dynamical simulations of the non-interacting model (see Fig.~\ref{nonint_fig}), we observe a sharp transition in time similar to Fig.~\ref{Z_vs_f}. Starting from an initial state without errors, the thermal environment introduces a growing number of spin-flips which eventually cause the memory to fail. This occurs again at a certain threshold probability $f_c$ which is for this case, however, different from $0.1$, see the inset of Fig.~\ref{Z_vs_f}. To understand this difference, we note that a main mechanism for the creation of errors is the diffusion of anyons. Clearly, errors created by the anyons in their diffusive motion have strong spatial correlations, rather than being independent and uniformly distributed across the memory. We find that such correlations yield values of $f_c$ strictly smaller than $0.1$ but still of the order of a few percent, see Fig.~\ref{Z_vs_f}. Although the value of $f_c$ is difficult to determine in general, we will assume in the following that such threshold probability exists and derive from it an expression for the memory lifetime. 

\begin{figure}
    \includegraphics[width=0.95\columnwidth]{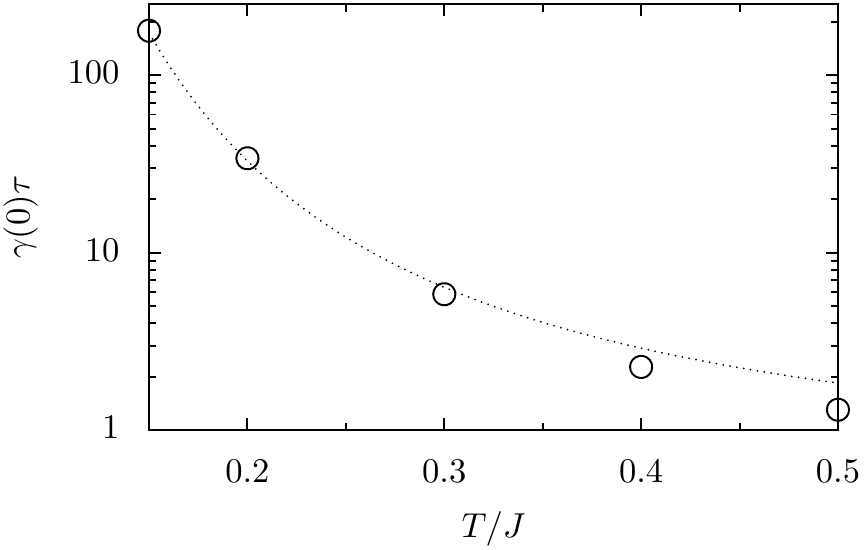}
    \caption{\label{nonint_T} The values of $\tau$ extracted at the sharp transitions of the $\langle Z_{\rm ec} \rangle$ decay (circles). As in Fig.~\ref{nonint_fig}, we use $\gamma(0)=\gamma(2J)$. Comparison to Eq.~(\ref{T_estimate}) (dotted curve) gives good agreement for $f_c\simeq 0.1$.}
\end{figure}

\subsection{Direct and indirect diffusion of anyons} \label{app:indirect}

To estimate the error creation rate, we first study the diffusive motion of anyons in the non-interacting model. To determine the diffusion constant $D$, we consider an isolated anyon in the lattice and its probability $p_{i,j}$ to be at site $(i,j)$. In the Ohmic case, we have $\gamma(0)\neq 0$, and direct hopping to neighboring sites is thus allowed. In the continuum limit, a standard diffusion equation $\frac{dp({\bf r})}{dt}=D\nabla^2 p({\bf r})$ with $D=\gamma(0)$ is obtained. The resulting decay of the bare and error-corrected logical operators in the simple case of a single pair is discussed in Appendix~\ref{app:single_pair}.

For a super-Ohmic bath where $\gamma(0)=0$, diffusion is still possible due to `indirect hopping'. We assume $2\beta J \gg 1$, such that, since $\gamma(2J)=e^{2\beta J}\gamma(-2J)$, the recombination of a pair of anyons is essentially instantaneous. Hopping from the site $(i, j)$ to, e.g., $(i,j+2)$ is possible by creation of an anyon pair occupying sites $(i,j+1)$ and $(i,j+2)$. This event occurs with rate $\gamma(-2J)$. Since the intermediate state can decay back to the initial state, the actual rate for the indirect hopping process is $\gamma(-2J)/2$. Similar considerations hold for all other sites. Accounting for all of these, we write
\begin{eqnarray}
\frac{dp_{i,j}}{dt}=\frac{\gamma(-2J)}{2}(-12 p_{i,j}+p_{i+2,j}+p_{i-2,j}+p_{i,j+2}+p_{i,j-2}  \nonumber  \\
+2p_{i+1,j+1}+2p_{i+1,j-1}+2p_{i-1,j+1}+2p_{i-1,j-1}), \nonumber
\end{eqnarray}
which, in the continuum limit, yields $D=4\gamma(-2J)$. We can expect
that the properties of the memory improve by lowering the value of
$\gamma(0)$, but only as long as $\gamma(0)\gtrsim 4\gamma(-2J)$. In
the interacting case, $J$ can be replaced by an appropriate excitation
energy (e.g., a mean-field gap, see Sec.~\ref{sec:mf}).

\subsection{Lifetime of the non-interacting model}\label{sec:lifetime_formula}

We can now express the error creation rate in terms of the diffusion constant. An isolated anyon can have either one or three $\sigma_x$-errors at its plaquette spins. In the first case, the anyon can hop to a neighboring site either by creating an error on one of the initially unaffected spins, or by removing the one pre-existing error. Therefore, such an anyon contributes to the error rate with $2D = 3D-D$. If three $\sigma_x$-errors are present, an opposite rate $-2D$ is obtained from an analogous reasoning. However, three-error plaquettes can be expected to be less likely: they require that an anyon hopped on a plaquette with two pre-existing errors from one of the two directions without errors. From the above discussion, it is justified to estimate the rate at which errors are created to be of order $D$ per anyon. 

Finally, assuming $N$ diffusing anyons present in the system, the fraction $f$ of spins affected by a $\sigma_{x}$-error after a time $t$ is estimated as $f \simeq N D t/2 L^2$ and the error correction fails when $f$ is larger than some critical value $f_c$ \cite{Dennis2002}. 
This gives a lifetime $\tau$ for the memory
\begin{equation}\label{T_estimate}
\tau \simeq 2 f_{c} \frac{e^{\beta J}+1}{\max\{ \gamma(0),4\gamma(-2J)\}},
\end{equation}
where we have replaced the factor $N/L^2$ by the equilibrium occupation $\langle n_p \rangle = 1/(e^{\beta J}+1)$. 

An analogous result can be obtained based on the following different reasoning \cite{Alicki2009, Hamma2009}. The distance between the two anyons of a pair after a time $\tau$ is of order $\Delta\ell=\sqrt{D \tau}$ and is required to be much smaller than the average anyon separation $\sim \sqrt{L^2/N}$. This gives 
$\tau \ll (e^{\beta J}+1)/\max \{ \gamma(0),4\gamma(-2J)\}$.
Interestingly, this upper bound coincides with the right-hand side of Eq.~(\ref{T_estimate}) if the probability for each spin to be flipped is $\frac12$ (which is realized at long times). 

Equation~(\ref{T_estimate}) generally gives reasonable estimates of the memory lifetime. For example, the value $f_c \simeq 0.11$ of \cite{Dennis2002} yields $\tau\simeq 5.8$ for the same parameters as used in Fig.~\ref{nonint_fig}, in remarkable agreement with the simulations. However, the real threshold directly obtained by the simulation is smaller (inset of Fig.~\ref{nonint_fig}). This seems not surprising considering the approximations introduced when deriving Eq.~(\ref{T_estimate}). We generally adopt the practice of using $f_c$ as a single fitting parameter to study the functional dependence of the lifetime, e.g., as a function of $L$ or $T$. An example of the temperature dependence of $\tau$ in the non-interacting case is shown in Fig.~\ref{nonint_T} and is also well described by Eq.~\eqref{T_estimate}.

More importantly, Eq.~(\ref{T_estimate}) allows one to describe the asymptotic dependence of the lifetime on $L$. For the non-interacting case, $\tau$ is independent of the system size, consistent with previous findings
\cite{Alicki2009,Nussinov2008,Hamma2009, Bravyi2009, Kay2008}. This fact is confirmed by our simulations, as shown in
Fig.~\ref{nonint_fig}, where $\langle Z_{\rm ec} \rangle$ clearly
approaches a step-function with increasing $L$. We also see that
the bare expectation value $\langle Z \rangle $ decays even faster with
larger $L$. Indeed, at sufficiently short times $t \ll 1/\max\{
\gamma(0),4\gamma(-2J)\}$, when anyon pairs have not yet diffused
apart from each other (the `nonsplit-pair' regime, indicated by an
asterisk), we obtain $\langle  Z \rangle= (1-
1/L)^{N^\ast/2} \simeq e^{-N^\ast/2L}$. By using $N^\ast \simeq 4
L^2 \gamma(-2J) t$, it follows that $\langle  Z \rangle$ decays
exponentially with $L$.

For the interacting case, we find good agreement of a modified version of Eq.~(\ref{T_estimate}) with the simulations [see Eq.~(\ref{T_estimate_int}) and Fig.~\ref{Zav_90thresh_int}]. Fitting the data always yields values of $f_c$ smaller than $f_c = 0.11$, but still of the order of a few percent. These values are thus consistent with the original meaning of $f_c$. For a more
extended discussion, we refer to Secs.~\ref{sec:lifetime_mf} and \ref{sec:lifetime_interacting}. 

\section{Mean-field analysis of the interacting model}\label{sec:mf}

We now turn to the interacting case $A > 0$ and perform a mean-field analysis, which becomes accurate in the relevant limit of large $L$. 

\subsection{Mean-field anyon density}

We first consider the equilibrium number of anyons $N$ within a mean-field treatment (mean-field values will be indexed with a subscript `mf'). 
We obtain the single-particle energy at plaquette $p$ as $\epsilon_p=\delta H_0/\delta n_p= J+ \sum_{p'\neq p} U_{p p'} n_{p'}$.
Replacing $n_{p'}$ by the average value $n_{\rm mf}=N_{\rm mf}/L^2$ and taking the continuum limit, we find the mean-field value for $\epsilon_p$ to be
\begin{equation}\label{Nmf}
\epsilon_{\rm mf}=
J+ n_{\rm mf}\int_{L \times L}  \frac{A}{r^\alpha} d{\bf r} =
J+ n_{\rm mf} T L_\alpha,
\end{equation}
where we use the notation 
\begin{equation}
L_{\alpha}=c_\alpha \beta A L^{2-\alpha}.
\end{equation} 
The constant $c_\alpha$ is a geometrical factor of order 1, given by the integration of $1/r^\alpha$ on a unit square centered at the origin. In particular, $c_0 = 1$. On the other hand, we have $n_{\rm mf}=1/(e^{\beta\epsilon_{\rm mf}}+1)$ since the occupation numbers $n_p$ can only assume the values $0$ or $1$.
By using Eq.~\eqref{Nmf} to calculate $n_{\rm mf}$, we find the
self-consistent equation
\begin{equation}\label{meanfield}
n_{\rm mf} = \frac{1}{e^{\beta J+ n_{\rm mf} L_\alpha }+1},
\end{equation}
with the following expansion at large $L_\alpha$
\begin{equation}\label{leadingNmf}
n_{\rm mf}=\frac{1}{L_\alpha}\left[\ln L_\alpha - \ln\ln L_\alpha
- \beta J+\dots \right].
\end{equation}
Higher order terms in the square brackets are small if $\ln L_{\alpha}\gg \beta J, |\ln \ln L_{\alpha}|$.
For fixed temperature $T$ and interaction strength $A$, these conditions are always satisfied at
sufficiently large $L$ since $L_\alpha \propto L^{2-\alpha}$.

\begin{figure}
    \includegraphics[width=0.95\columnwidth]{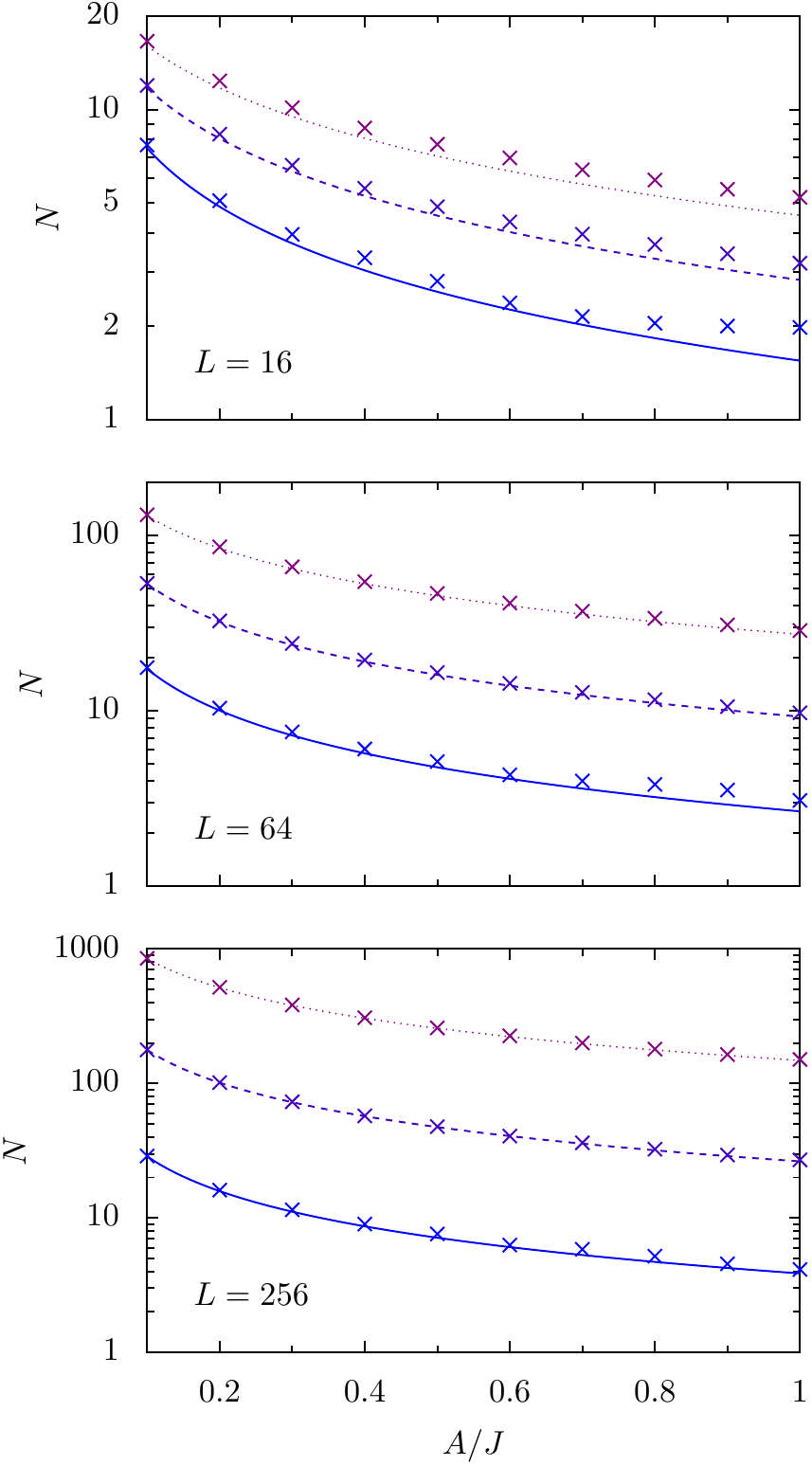}
    \caption{\label{N_metropolis} (Color online.) Comparison of the equilibrium value of $N$ obtained numerically (crosses) with $N_{\rm mf}$ (curves) for different grid sizes. We have used the interaction exponents $\alpha = 0$ (solid line), $\alpha = 0.5$ (dashed line), and $\alpha = 1.0$ (dotted line), and the temperature $T/J = 0.5$.}
\end{figure}

We have confirmed the validity of the mean-field approximation by Monte Carlo simulations. By using the Metropolis algorithm \cite{Metropolis1953} to sample the probability distribution
$\propto e^{-\beta/2 \sum{p,p'}U_{pp'}n_p n_{p'}}$, see
Eq.~(\ref{H0}), the equilibrium number of excited plaquettes can be approximated
with arbitrary accuracy. This can be used to study the accuracy of the mean-field
value $N_{\rm mf}=n_{\rm mf}L^2$ [see Eq.~(\ref{meanfield})], in particular for values $\alpha \neq 0$.
Due to the long-range nature of the interaction, $N_{\rm mf}$
compares very well to the equilibrium value of $N$ obtained from these simulations at generic
values of the temperature and interaction exponent $\alpha$. This is
illustrated in Fig.~\ref{N_metropolis}, which further shows a satisfactory
agreement already at moderate values of $L$.

We also note that, for the case of constant interaction ($\alpha=0$), the average number can be calculated directly from the grand-canonical partition function
\begin{equation}\label{part_funct}
{\sum_{2k\leq L^2}}\left (
\begin{array}{c}
L^2 \\
2k
\end{array}
\right) e^{-\beta E_{2k}},
\end{equation}
since the energy of a given anyon configuration does not depend on the positions of the anyons, but only on their total number $N=\sum_p n_p$. In the presence of a sufficiently strong anyon interaction or at low temperature, the number of excited plaquettes is much smaller than $L^2$. Therefore, one can restrict the sum (\ref{part_funct}) to the first few relevant terms.

\subsection{Lifetime of the interacting model}\label{sec:lifetime_mf}

From Eq.~(\ref{leadingNmf}) we obtain that, even though the number of anyons $N_{\rm mf}$ grows with the system size $L$, the anyon density $n_{\rm mf}$ goes to zero for long-range repulsive interactions with $0 \leq \alpha < 2$. Hence, the population of anyons is increasingly diluted and the system is essentially frozen in the ground state at large system size. This remarkable effect can be attributed to the divergence of the excitation energy $\epsilon_{\rm mf}\simeq T \ln L_\alpha$, which is self-consistently determined from the anyon population in the \emph{whole} system due to the long-range nature of the interactions. Note also that, despite the fact that $\epsilon_{\rm mf}$ is diverging, the total excitation energy density $n_{\rm mf}\epsilon_{\rm mf}/2$ goes to zero for large $L$.

Secondly, the divergence of $\epsilon_{\rm mf}$ leads to a vanishing anyon pair creation rate at large $L$,
\begin{equation}\label{gamma_2emf}
\gamma(- 2\epsilon_{\rm mf}) \simeq \kappa_n T^n \frac{(2\ln L_\alpha)^{n+2}}{2 L_\alpha^2}.
\end{equation}
This fact allows us to revise the lifetime for the non-interacting memory Eq.~(\ref{T_estimate}), simply by substituting $J$ with the equilibrium value $\epsilon_{\rm mf}$, yielding
\begin{equation}\label{T_estimate_int}
\tau \simeq \frac{2 f_{c} / n_{\rm mf}}{\max\{ \gamma(0),4\gamma(-2\epsilon_{\rm mf})\}}.
\end{equation}
From this we obtain the lifetime of an interacting memory in case of an Ohmic ($n=1$) or super-Ohmic ($n > 1$) bath as
\begin{equation}\label{T_estimate_int1}
\tau \simeq
\begin{cases} 
\dfrac{f_c L_\alpha}{\kappa_1 T \ln L_\alpha} , & \text{Ohmic}
\vspace{0.3cm}
\\
\dfrac{2 f_{c}L_\alpha^3}{\kappa_n T^n (2\ln L_\alpha)^{n+3}} , &\text{super-Ohmic}
\end{cases}
\end{equation}
in the limit of large grid size [see after Eq.~(\ref{leadingNmf})]. It is clear from these expressions that the memory lifetime is diverging with $L$, in strong contrast to the non-interacting case where it was bounded by a constant. In the Ohmic case, this divergence of $\tau$ is entirely due to the vanishing density, since $\gamma(0)= 2 \kappa_1 T$ is non-zero. In the super-Ohmic case, however, an additional divergence due to the vanishing of $\gamma(-2\epsilon_{\rm mf})$ is obtained, see Eq.~(\ref{gamma_2emf}). Since the energy gap grows logarithmically with $L$, $\tau$ grows polynomially, but with a rather favorable power.
For instance, constant interaction ($\alpha=0$, see also below) leads to $\tau \propto L^2/\ln L$ in the Ohmic case
and to $\tau \propto L^6/\ln^5 L$ in the super-Ohmic ($n=2$) case.

\subsection{Effects beyond the mean-field treatment}

Equation~(\ref{T_estimate_int}) is valid in the mean-field limit and does not include effects of the fluctuations of the number of anyons and their positions. These result in additional errors and correlated spin-flips across the memory, due to the long-range nature of the anyon interactions. Although we expect in general deviations from Eq.~(\ref{T_estimate_int}), the memory remains self-correcting both for an Ohmic and for a super-Ohmic bath. 

Indeed, for an Ohmic bath, we can neglect the effect of the repulsive force if the change of energy $\omega$ in a diffusive step is smaller in magnitude than $T$ [see Eq.~(\ref{gamma_bath_text})], so that we can approximate $\gamma(\omega)\simeq \gamma(0)$. In particular, for a single pair of anyons at distance $r$, we have $|\omega| \lesssim \alpha A/r^{\alpha+1} $, which defines a critical radius
\begin{equation}
r_c =( \alpha A\beta)^{\frac{1}{\alpha+1}},
\end{equation}
beyond which the fluctuations become negligible. For $\alpha=0$ one has $r_c=0$. For $\alpha >0$, since the average distance $\sim 1/\sqrt{n_{\rm mf}}$ between anyons grows with $L$ while $r_c$ is independent of $L$, the fluctuations also become negligible. The validity of Eq.~(\ref{T_estimate_int}) for the Ohmic case is confirmed by numerical simulations both for $\alpha=0$ (see Fig.~\ref{Zav_90thresh_int}) and for $\alpha>0$ (see Fig.~\ref{non_constant_interaction}).

Concerning the super-Ohmic case, Eq.~(\ref{T_estimate_int}) could become inaccurate if the fluctuations of $\omega \simeq 0$ are more effective for the anyon motion than the indirect diffusion mechanism which is proportional to the rate in Eq.~(\ref{gamma_2emf}). However, due to the decreasing interaction strength, such fluctuations in $\omega$ become small at large $L$ and still result in a vanishing diffusion coefficient. Therefore, Eq.~(\ref{T_estimate_int}) might overestimate the lifetime in this case, but the asymptotic dependence on $L$ would still be better than in the Ohmic case. Furthermore,  at $\alpha=0$ direct hopping is impossible and Eq.~(\ref{T_estimate_int}) is valid (see Fig.~\ref{Zav_90thresh_int}). 
 
\section{Dynamics of the interacting model}\label{sec:lifetime_interacting}

\begin{figure}
    \includegraphics[width=0.95\columnwidth]{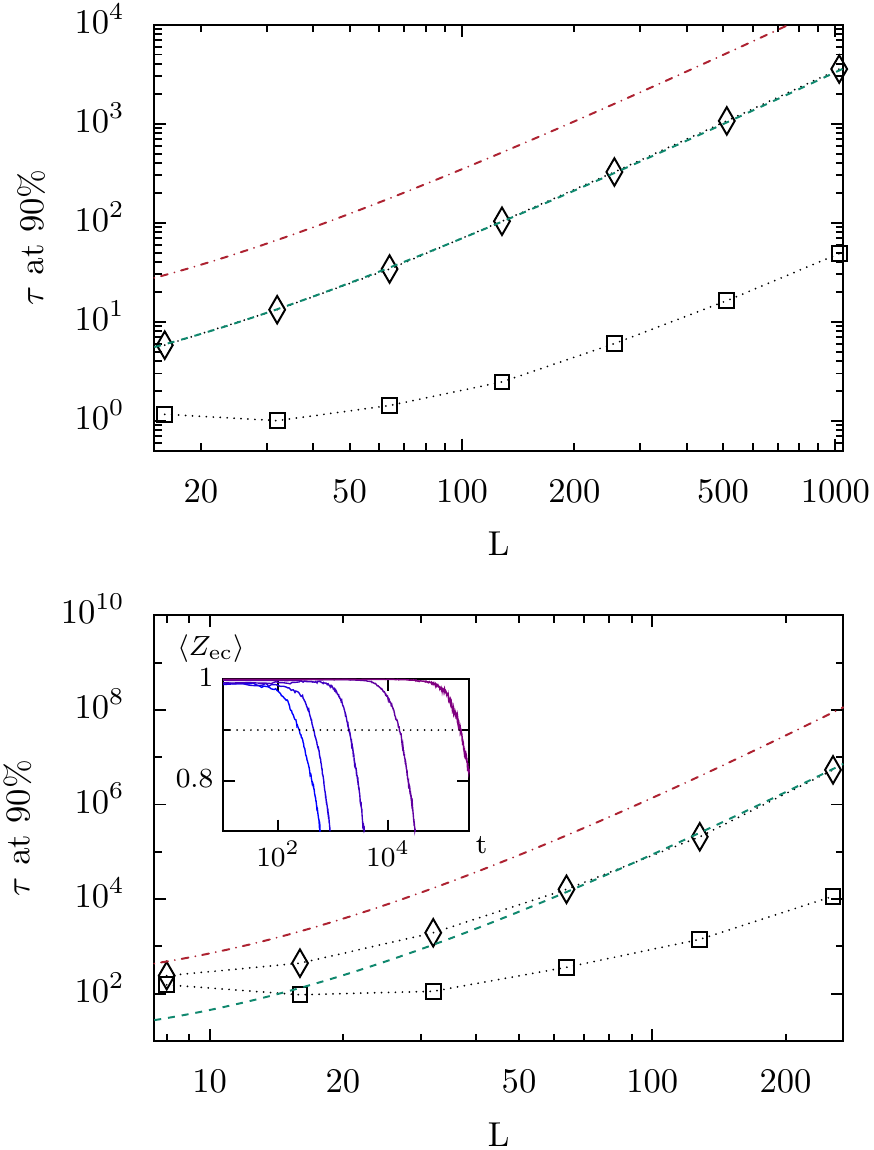}
    \caption{\label{Zav_90thresh_int} (Color online.) Thermal stability of the interacting memory. The data in the top (bottom) panel were obtained for an Ohmic (super-Ohmic, $n = 2$) bath. Plotted as a function of $L$ are the numerically simulated times at which the expectation values of the bare (squares) and error-corrected (diamonds) logical $Z$ operator have decayed from $1$ to $0.9$.
    The dotted lines serve as a guide to the eye. The red dashed-dotted curves are calculated from Eq. \eqref{T_estimate_int} with $f_c = 0.11$, where we have used the self-consistent values of $n_{\rm mf}$ and $\epsilon_{\rm mf}$ from Eqs.~(\ref{Nmf}) and (\ref{meanfield}). Similarly, the green dashed lines are also due to Eq. \eqref{T_estimate_int}, but here $f_c$ is fit to the numerical data of the $90\%$ threshold times, yielding $f_c = 0.022$ for an Ohmic, and $f_c = 0.007$ for a super-Ohmic bath. The inset shows the decay of $\langle Z_{\rm ec}\rangle$ with time for $L=8, \ldots, 128$ (from left to right), and the $90\%$ threshold is illustrated by the dotted line. It is seen that choosing this particular value has no substantial influence on the scaling behavior with $L$. Parameters used in these simulations were $A/J = 0.1$, and $T/J = 0.3$. Times are in units of $(\kappa_1 J)^{-1}$ and $(\kappa_2 J^2)^{-1}$ for the first and second panel, respectively. }
\end{figure}

We turn now to the numerical simulations of our model, Eq.~(\ref{H0}), and focus first on constant long-range interactions ($\alpha=0$). In this case, the total energy $E_N= N J +\frac{A}{2}N(N-1)$ depends only on the number of anyons $N$, but not on their position. This
simplifies the numerical treatment considerably.
Our results are displayed in Fig.~\ref{Zav_90thresh_int}. The numerical data show a clear increase of the memory lifetime $\tau$ with $L$. Note that this holds already for the \emph{bare} logical $Z$ operator.
Like in the non-interacting case (see Fig.~\ref{nonint_fig}), the beneficial effect of the error correction at read-out is to prolong
 the lifetime by maintaining $\langle Z_{\rm ec} \rangle$ close to 1 (see inset of Fig.~\ref{Zav_90thresh_int}).
 
Our analytical results describe the numerical data remarkably well. By fitting $f_c$ in Eq. \eqref{T_estimate_int} to the simulation data, excellent agreement is found for an Ohmic bath (top panel of Fig.~\ref{Zav_90thresh_int}), while for a super-Ohmic bath (lower panel), analytics and numerics agree well for $L \gtrsim 64$. 
Furthermore, the fit yields values for $f_c$ of about $0.01 - 0.02$, which is
reasonable in comparison to the upper bound $f_c = 0.11$ found for a model of uncorrelated errors (dashed-dotted lines in Fig.~\ref{Zav_90thresh_int}) \cite{Dennis2002}. See also the discussion in Sec.~\ref{sec:lifetime_formula}. 

The lifetime $\tau$ can be compared to the physical time scales of single spin flips, $1/\gamma(0)$ and $1/\gamma(-2J)$.
For instance, for the $L=256$ super-Ohmic case in Fig. 3 we obtain
$\tau \gamma(-2J)\simeq 5\times 10^5 $, i.e., already for a moderate 
system size the lifetime $\tau$ of the memory is
about a $10^6$ times longer than the single-spin lifetime. 
For quantum dots, the latter is typically in
the range of milliseconds to seconds at about $100\,{\rm mK}$ \cite{Golovach2004, Amasha2008}.

\subsection{The nonsplit pair regime}

We consider now in greater detail the super-Ohmic case at $\alpha=0$, which has the most favorable scaling. The initial dynamics of the memory can be nicely characterized by a regime of nonsplit pairs. Under this assumption, the rate equation
\begin{equation}\label{rate_eq_pairs_int}
\frac{dN^*_{\rm mf}}{dt}=4 L^2 \gamma(-2\epsilon^*_{\rm mf})-N^*_{\rm mf} \gamma(2\epsilon^*_{\rm mf})
\end{equation}
describes the initial time-evolution of the system well, since in this non-diffusive regime only pair creation \cite{footnote02} and annihilation takes place. In Eq.~(\ref{rate_eq_pairs_int}) we denote with $N_{\rm mf}^*$ the total number of anyons, appearing as $N_{\rm mf}^*/2$ nonsplit pairs. 

\begin{figure}
    \includegraphics[width=0.95\columnwidth]{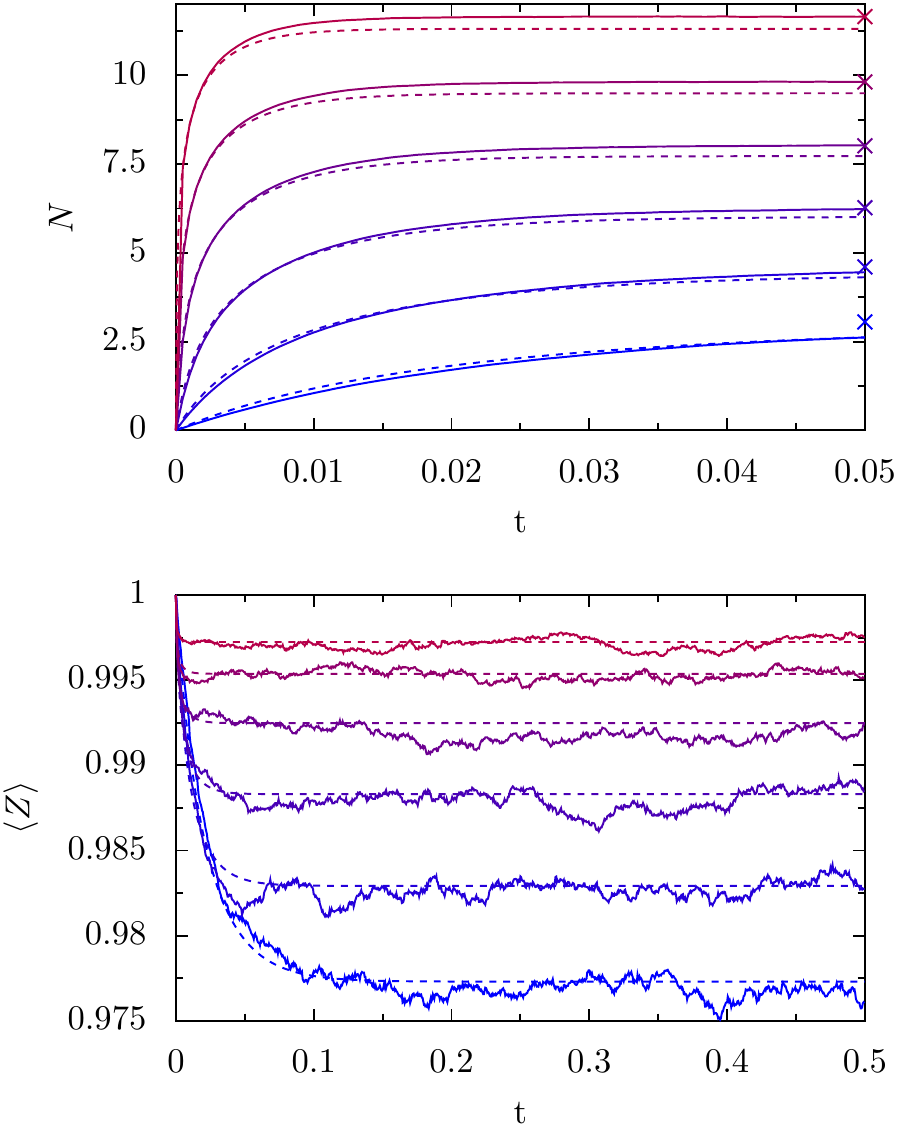}
    \caption{\label{pair_creation} (Color online.) Short-time dynamics of the interacting memory in a super-Ohmic bath.
    In this case, the memory is in the nonsplit-pair regime. The curves refer to different values of $L$ increasing in powers of 2 from $L = 64$ (lowest curves in both panels) to $L = 2048$ (highest curves).
    Upper panel: The time dependence of the anyon
    number $N$ obtained from the simulations (solid lines) is compared to the solutions
    of Eq.~(\ref{rate_eq_pairs_int}) (dashed lines).
    The crosses are the exact values $N^*$ obtained from the partition
    function of pairs  Eq.~(\ref{Z_pairs}). Good agreement with $N^*$ is also obtained for the lower curves at
    longer times (not shown).
    Lower panel: The expectation value of the bare $Z$ obtained from the simulations
    (solid lines) is compared to $e^{-N^*/2L}$ (dashed lines), where $N^\ast(t)$ is obtained from the upper plot.
    Parameters used are $A/J = 0.1$, and $T/J = 0.3$. The time axes are in units of $(\kappa_2 J^2)^{-1}$.}
\end{figure}

We confirm Eq.~\eqref{rate_eq_pairs_int} by comparing its solution, obtained by numerical integration, with a direct simulation presented in Fig.~\ref{pair_creation}. After a rapid initial `build-up' phase, $N_{\rm mf}^\ast$ saturates to a value determined by the self-consistent condition $N_{\rm mf}^{*} = 4 L^2 e^{-2(J+AN_{\rm mf}^{*})\beta}$, obtained by setting $dN^\ast_{\rm mf}/dt = 0$ in Eq. \eqref{rate_eq_pairs_int}. In this state, the excitation energy is diverging with $L$, since we have $\epsilon^\ast_{\rm mf} \simeq AN^\ast_{\rm mf} \simeq AN_{\rm mf}/2 \propto \ln L$. This effectively suppresses the indirect diffusion of anyons. Therefore, the system remains in a quasi-stationary state which evolves to the final anyon density on a time scale also diverging with $L$.
In this regime of nonsplit pairs, one has $\langle Z \rangle \simeq e^{-N^\ast_{\rm mf}/2L}$. This leads to the quasi-stationary value $\langle Z \rangle \simeq e^{-\frac{\ln L}{2 \beta A L}}$, which approaches 1 for large $L$ (see Fig.~\ref{pair_creation}).

Similar to the calculation of the total number of anyons [see Eq.~(\ref{part_funct})], the exact quasi-stationary number of paired anyons $N^*$ (crosses in Fig.  Fig.~\ref{pair_creation}) can be calculated from a partition function reading
\begin{equation}\label{Z_pairs}
{\sum_{k\leq 2L^2}} \left(
\begin{array}{c}
2L^2 \\
k
\end{array}
\right)e^{-\beta E_{2k}}~.
\end{equation}
Here we have assumed that $k$ sufficiently diluted errors are present in the memory such that $2k$ anyons are created in the nonsplit-pair regime. The average number of anyons $N^*$ calculated from Eq.~(\ref{Z_pairs}) is in very good agreement with the simulations, see Fig.~\ref{pair_creation}.

\subsection{Non-constant interaction}

\begin{figure}
    \includegraphics[width=0.95\columnwidth]{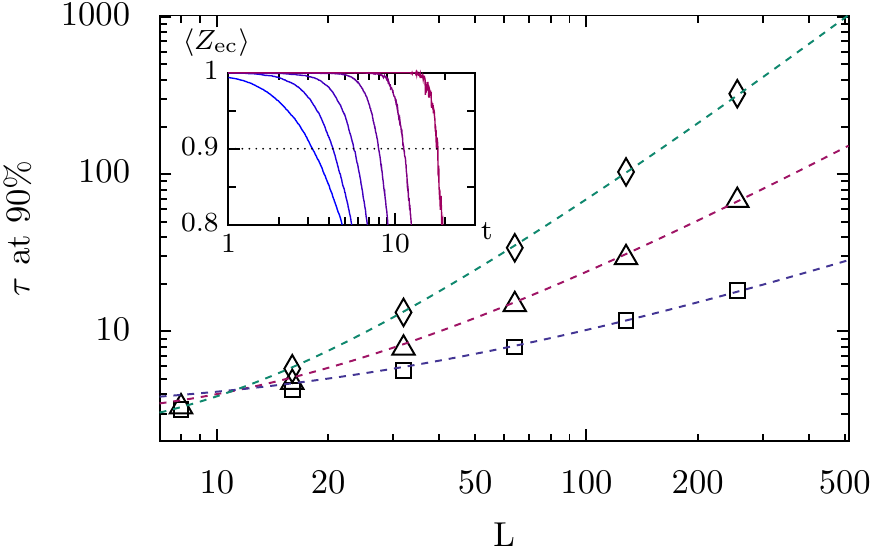}
    \caption{\label{non_constant_interaction} {(Color online.) Thermal stability of the interacting memory with $\alpha \neq 0$ and an Ohmic bath.} Data points refer to the numerically calculated times at which the error-corrected logical $Z$ operator has decayed from $1$ to $0.9$ in the cases $\alpha = 0$ (diamonds), $\alpha = 0.5$ (triangles), and $\alpha = 1$ (squares). Note that we have replotted the data from $\alpha = 0$ merely for comparison. The dashed lines are from Eq.~\eqref{T_estimate_int} (as in Fig.~\ref{Zav_90thresh_int}), with a fit of $f_c$ yielding $f_c = 0.027$ for $\alpha = 0.5$ and $f_c = 0.032$ for $\alpha = 1$. Inset: Decay of $\langle Z_{\rm ec}\rangle$ as a function of time for different grid sizes, $L = 8, 16, \ldots, 256$ (left to right), and $\alpha = 1$. Parameters used in the simulations were $A/J = 0.1$, and $T/J = 0.3$. Times are in units of $(\kappa_1 J)^{-1}$.}
\end{figure}

For non-constant long-range interaction ($0<\alpha<2$), simulating the time dynamics of the memory is numerically more costly due to an $O(L^2)$ overhead coming from recalculating all spin flip rates. Nevertheless, we were able to study the (more tractable) case of an Ohmic bath. The results are presented in Fig.~\ref{non_constant_interaction} for $\alpha=0.5$ and $\alpha=1$. Clearly, the memory lifetime is still increasing with $L$, proving the memory to be self-correcting also for $\alpha \neq 0$. Furthermore, the data are in very good agreement with the analytically calculated lifetime Eq.~(\ref{T_estimate_int}). The super-Ohmic case for $\alpha>0$ is more difficult to simulate due to the increased memory lifetime and will be examined elsewhere.

\section{Discussion of the long range interaction}\label{sec:long_range}

So far we have assumed the presence of long-range anyon interactions. We briefly comment here on their possible realization. Concerning the many-body nature of the interactions involved,
 general $n$-body couplings can in principle be engineered from two-body interactions \cite{Kempe2006, Bravyi2008, Wolf2008, Jordan2008}. For example, toric codes with interacting anyons are derived in \cite{Hamma2009, Schmidt2008}.
 A systematic procedure to construct such effective low-energy Hamiltonians can be rigorously founded on the Schrieffer-Wolff transformation \cite{Bravyi2008, Wolf2008}. In a similar way, physical long-range interactions of the type considered in this work could also be generated perturbatively. A well-known example is
the Ruderman-Kittel-Kasuya-Yosida (RKKY) interaction \cite{Kittel1987}, e.g., for a 2D Kondo-lattice of nuclear spins \cite{Simon2007}. Alternatively, constant interactions ($\alpha=0$) can be realized for qubits coupled to photon modes in QED-cavities \cite{Dicke1954, Pellizzari1995, Imamoglu1999, Wallraff2004, Burkard2006, Trif2008}. The interaction range is determined by the wavelength of the photon and can reach macroscopic distances, in particular in superconducting cavity striplines \cite{Wallraff2004, Burkard2006, Trif2008}. Another promising candidate system to realize topological models are ultracold atoms or molecules in optical lattices \cite{Jiang2008,Weimer2010}. 

As a most elementary example, consider all plaquette operators interacting with a delocalized two-level system (acting as an ancilla qubit), in analogy to the so-called central spin problem. For example, $H_{\rm int}= \Delta \sigma_z + \sum_p g_p n_p \sigma_x$ with eigenvalues  $\pm  \sqrt{\Delta^2 + (\sum_p g_p n_p)^2}$. A quadratic expansion of the higher eigenvalue $\simeq \Delta + \frac{1}{2\Delta} \left(\sum_p g_p n_p \right)^2 $ (if $\Delta>0$) gives a repulsive interaction between the anyons. Note that in this example the central spin has to be kept in the excited state.  

A physically more interesting case is the two-photon coupling described by the Hamiltonian
\begin{equation}\label{eq:two-photon coupling}
H_{\rm int} = \sum_{i = 1}^2 \omega_i a_i^\dag a_i + \sum_p g_p n_p (a_1^\dag a_2 + a_1 a_2^\dag).
\end{equation}
Here, $\omega_i$ are the photon frequencies, and $g_p$ is the coupling strength of plaquette $p$ to the modes. This type of coupling naturally emerges in the perturbative derivation of the toric code model from the Kitaev honeycomb model \cite{Kitaev2006} if a quadratic coupling to electric (or magnetic) cavity fields such as $E_xE_y$ is added. We start from the expression of the anyon excitation energy obtained in leading order of perturbation theory, given by
\begin{equation}\label{J0_from_Jk}
J_0 = \frac{J_{x}^2J_{y}^2}{8J_{z}^3},
\end{equation}
where $J_k$ are the exchange couplings in the honeycomb lattice \cite{Kitaev2006,Schmidt2008}. Since the couplings $J_k$ are determined by exchange integrals, they can be modified by electric perturbations: In multiferroic materials, electric fields can couple to the spin (-texture) via a modification of the exchange interaction such as
$J_k\rightarrow J_k+\delta_k (a_k + a_k^\dag)$ \cite{Trif2008a,Trif2008b} (with $\delta_{x,y,z}$ being some coupling constants and $a_{x,y} \equiv a_{1,2}$). Thus, if, for example, one $J_x$ and one $J_y$ occurring in $ J_{x}^2J_{y}^2/8J_{z}^3$ get modified in this way (by locally modifying the corresponding links), we end up with a coupling of the desired form with
\begin{equation}\label{g_from_Jk}
g_p = \frac{J_x J_y \delta_x \delta_y}{2J_z^3}.
\end{equation}
A possible concern is that the spin-electric couplings introduce several other interaction terms in addition to Eq.~(\ref{eq:two-photon coupling}) \cite{comment_stars}. By imposing the resonance condition $\omega_1 \approx \omega_2$, the quadratic term $(a_1^\dag a_2 + a_2^\dag a_1)$ can be made dominant over the linear ones (which are non-resonant). Furthermore, higher-order terms can be neglected for $\delta_k \ll J_k$ (a more detailed analysis will be presented elsewhere \cite{PedrocchiUnpublished}).

The Hamiltonian Eq.~\eqref{eq:two-photon coupling} can be brought to the diagonal form $H_{\rm int}=\sum_{i=1}^2 \Omega_i b_i^\dag b_i$ by making use of a standard Bogoliubov transformation of the boson operators. Since $g_p$  is spatially constant over the photon wavelength $\lambda_{i}$ \cite{Dicke1954}, we assume in the following a constant value $g_p=g$, such that $\sum_p g_p n_p = g N$. Therefore,
\begin{eqnarray}
b_1=\cos\theta a_1 + \sin\theta a_2,\\
b_2=\cos\theta a_2 - \sin\theta a_1,
\end{eqnarray}
with $\tan 2\theta = 2g N/(\omega_1-\omega_2)$ and
\begin{equation}\label{eq:Omegas}
\Omega_{1,2}=\frac{\omega_1+\omega_2}{2} \pm \sqrt{\left(\frac{\omega_1-\omega_2}{2}\right)^2+ \left(g N \right)^2}.
\end{equation}
By expanding $H_{\rm int}$ to lowest order in $g$ we obtain the desired constant anyon interaction,
\begin{equation}\label{eq:effective interaction hamiltonian}
H_{\rm int} \simeq  \sum_{i = 1}^2 \omega_i b_i^\dag b_i + \frac{b_1^\dag b_1 - b_2^\dag b_2}{\omega_1 - \omega_2}\left(g N \right)^2.
\end{equation}
The same result can also be derived with the general method of the Schrieffer-Wolff transformation \cite{Imamoglu1999,Trif2008} (see also Appendix~\ref{app:SW_trafo}). The strength and sign of the interaction  are tunable via the difference in frequencies and occupation numbers of the modes, and can consequently be made repulsive in a steady-state regime. We identify the parameters of Eqs.~(\ref{H0}) and (\ref{Upp_long_range}) as follows
\begin{equation}\label{Anew}
J=J_0 +  \frac{g^2}{\omega_1 - \omega_2}\langle b_1^\dag b_1  \rangle
\quad {\rm and} \quad 
A=  \frac{2 g^2}{\omega_1 - \omega_2}\langle b_1^\dag b_1 \rangle.
\end{equation}
The value of $J$ includes a small self-energy correction. For definiteness, we assumed that only the first mode (with $\omega_1> \omega_2$) is populated while $\langle b_2^\dag b_2  \rangle=0$.

Similarly to the first example, the case of repulsive interaction corresponds to a larger occupation of the mode with higher frequency. This condition is never realized in equilibrium and thus requires excitation of the cavity mode, which is easily accomplished by an external laser. Therefore, this specific realization of the long-range interaction corresponds to some sort of optical pumping of the memory into its ground state. It allows to avoid the full machinery of active error-correction, but cannot be considered passive in the strict sense of the term. 

Finally, while a non-equilibrium regime is generally needed for interactions obtained in second-order perturbation theory, it might be possible to derive repulsive interactions in the ground state at higher orders by a more elaborate construction. 


\section{Conclusion}\label{sec:conclusion}

We have discussed a generalization of the Kitaev toric code to include repulsive 
long-range anyon interactions. The properties of the system have been analyzed within a mean-field treatment, 
which we find to become accurate at large system size. Additionally, we have numerically studied the system dynamics via direct simulations. This has allowed us to 
demonstrate robust storage of the information encoded in the ground state manifold at large 
system size.

A similar model to ours, but with attractive instead of repulsive long-range interactions, was studied in Ref.~\cite{Hamma2009}, and was also found to
possess self-correcting properties. In that case, however, the interaction is 
logarithmically divergent with distance while we consider here more physical interactions, i.e.,
polynomially decaying. A dependence of this type is commonly found in condensed matter systems and, 
more specifically, we show that local coupling of the anyon operators to long-range optical modes would allow to
realize such interactions. As for the periodic boundary conditions, these are not an essential 
ingredient to a topological stabilizer code~\cite{Bravyi1998, Freedman1998}.

Another important aspect of our study is that the properties of the memory are strongly influenced by the
type of thermal bath. We obtained the size dependence of the memory lifetime for Ohmic
and super-Ohmic baths, the latter representing an especially advantageous situation. For example, 
for typical stripline cavities with $\lambda_{i} \sim $ cm and typical lattice constants of 100 nm (e.g. quantum dots), 
we see that the anyon interaction stays constant over system sizes $L$ as large as $10^{5}$. Extrapolating the 
super-Ohmic curve of Fig.~\ref{Zav_90thresh_int}, an enhancement factor $\sim 10^{20}$
is obtained at this value of $L$. With a single-spin lifetime $1/\gamma(-2J )\sim 1 \mu s-1 s$ \cite{Hanson2007, Amasha2008} 
this gives a memory lifetime $\tau \sim 10^{14}-10^{20}$~s. However, the assumption that the super-Ohmic scaling 
is valid up to this large size might be violated (e.g., because $\gamma(0)=0$ can only 
hold approximately).

In conclusion, we have demonstrated the existence of 2D stabilizer quantum memories at finite temperatures.
In our model, the stability of the memory is due to a large effective gap 
created by the repulsive interactions, which results in a vanishing anyon density.
Furthermore, the diffusive motion of the anyons is quenched in a super-Ohmic bath, when
the diffusion process requires creation of new anyon pairs. We expect that
similar systems in the presence of such interactions also prove useful 
as self-correcting quantum memories.

\acknowledgments

We would like to thank D. P. DiVincenzo, A. Imamoglu, and B. M. Terhal for discussions. This work was partially supported by the Swiss NSF, NCCR Nanoscience Basel, and 
DARPA.




\appendix

\section{Mapping from lattice gas to Ising model} \label{app:ising_chain}

Note that $H_0$ in Eq.~(\ref{H0}) has the general form of two independent lattice gases, which are in turn equivalent to two Ising spin lattices. We explicitly perform the transformation in the plaquette sector by identifying the Ising variables $s_p \equiv 1-2n_p$, yielding
\begin{equation}
H_0= -\sum_p\Big(\frac{J}{2}+{\sum_{p'}}^\prime\frac{U_{pp'}}{4}\Big)s_p +\frac18 {\sum_{p,p'}}^\prime U_{pp'} s_p s_{p'}+\ldots~,
\end{equation}
where $U_{pp'}$ is given in Eq.~\eqref{Upp_long_range} and the primes in the summations indicate $p'\neq p$. We have used $U_{pp}=2 J$ and $U_{pp'}=U_{p'p}$. The noninteracting Kitaev model corresponds to noninteracting spins in an external magnetic field. The ground state corresponds to the fully polarized state $s_p=1$ for all $p$, where no anyon is present. However, for $T>0$ a finite density of anyons emerges and is sufficient to destroy the information stored in the memory.

If a short-range ferromagnetic interaction is introduced, ordering of the system is spontaneously favored below some critical temperature. A higher magnetization corresponds to a lower population of anyons and improves the lifetime. However, short range interactions do not improve the scaling of the lifetime with the system size, since a residual density of anyons is left at any finite temperature. As in the noninteracting case, a finite density of excited plaquettes efficiently destroys the stored quantum information, in agreement with the general analysis of \cite{Bravyi2009, Kay2008}. Instead, repulsive long-range
interactions lead to a fully polarized system at a given temperature
for sufficiently large system size $L$.

\section{Lifetime in the presence of a single pair}\label{app:single_pair}

The decay of the bare and logical $Z$ operators is most simply
illustrated by assuming only a single anyon pair in the memory. We set
$\gamma(2J)=0$, so that pair creation and annihilation are not
allowed. If no anyons were present, the initial values $\langle Z
\rangle= \langle Z_{\rm ec} \rangle =1$ would be stable. We apply
one $\sigma_x$-operation at a randomly chosen site and thereby
create two neighboring anyons at $t=0$. This causes a partial decay
of the bare logical operator already at $t=0$, since we might have
chosen to flip a spin on the logical $Z$ operator, yielding $\langle
Z \rangle = 1-\frac{1}{L}$. This has been used in the main text in
the discussion of the nonsplit-pair regime.

\begin{figure}
    \includegraphics[width=0.95\columnwidth]{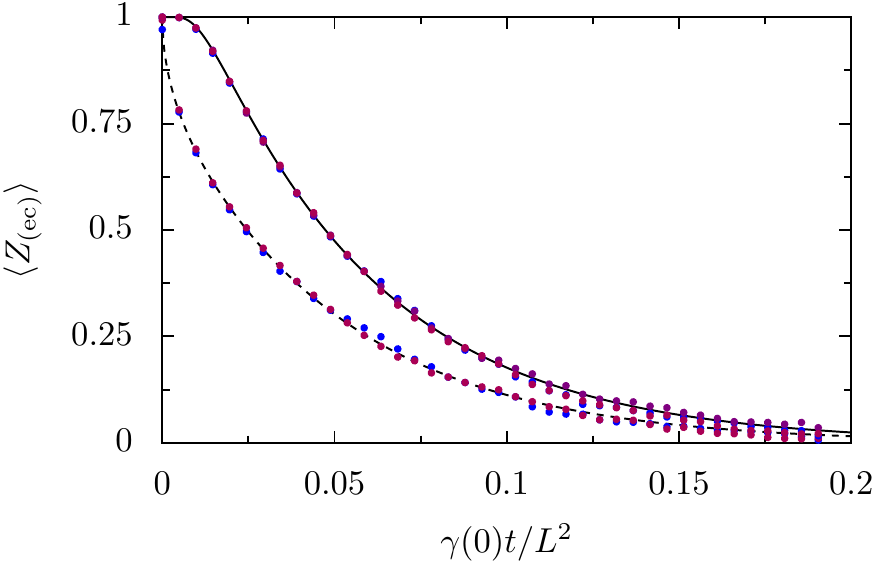}
    \caption{\label{pair_of_anyons} (Color online.) Decay of the bare and corrected expectation value of $Z$ due to a single pair of anyons in the memory. The dots show numerical data (averaged over $10^4$ samples) while the two curves are the continuum limit expressions Eq.~(\ref{P0_pair}) and (\ref{P_pair}) for $\langle Z_{\rm ec} \rangle$ (solid) and $\langle Z \rangle$ (dashed). The numerical data have been obtained for $L = 32, 64, 128$. All points collapse onto each other when plotted as a function of $\gamma(0)t/L^2$.}
\end{figure}

We now study the decay for $t > 0$ in the continuum limit and
therefore neglect the $1/L$ correction at $t=0$. We consider a single pair of diffusing anyons with
coordinates $(x_1, y_1)$ and $(x_2, y_2)$ created at
the origin. We then assume that the probability to find an anyon at position~$\bf r$ is described by the probability density
\begin{equation}
p({\bf r})=\frac{1}{4\pi \gamma(0) t} e^{-\frac{r^2}{4\gamma(0)t}}.
\end{equation}
We represent the torus as an infinite plane with the points $(x,y)$
and $(x+m L, y+n L)$ being equivalent ($m ,n\in \mathbb{Z}$). The
logical $Z$ operator is then represented by parallel lines at
$y_Z=L/2+n L$. The two anyons diffuse along $y$ with probability
density $p(y_i-y_0)=e^{-(y_i-y_0)^2/4\gamma(0)t}/\sqrt{4\pi
\gamma(0) t}$, where $i=1,2$ and the initial (random) coordinate
satisfies $-L/2\leq y_{0} < L/2$. The average of the logical
operator at time $t$ is
\begin{equation}
\langle Z \rangle=\int_{-L/2}^{L/2}\frac{dy_0}{L} \int dy_1 dy_2 p(y_1-y_0)p(y_2-y_0) z(y_1,y_2),
\end{equation}
where $z(y_1,y_2)$ gives the sign of $Z$ if the two anyons have
diffused to the coordinates $y_1$ and $y_2$. Since $Z$ changes sign
each time an anyon crosses the lines at $y_Z$, we have
$z(y_1,y_2)=z(y_1)z(y_2)$ where $z(y)=1$ if $-L/2+2nL \leq y
<L/2+2nL$ and $-1$ otherwise ($n\in \mathbb{Z}$). Therefore we can
write
\begin{equation}\label{P0_pair}
\langle Z \rangle =\int_{-1/2}^{1/2} dz_0 f(z_0)^2,
\end{equation}
where we have made the change of variables $y_0=L z_0$, such that
\begin{eqnarray}
f(z_0)= \frac12 \sum_{n=-\infty}^{+\infty}(-1)^n \left[{\rm erf}\left(\frac{2 z_0+2n+1}{4\sqrt{\gamma(0)t/L^2}}\right)  \right. \quad \nonumber \\
-\left. {\rm erf}\left(\frac{2 z_0+2n-1}{4\sqrt{\gamma(0)t/L^2}}\right)\right].
\end{eqnarray}

We now consider the average of the error-corrected logical operator $Z_{\rm ec}$. In this case, only the distance $y_{12}=y_1-y_2$ between the two anyons is important since the value of $Z_{\rm ec}$ is $1$ if $-L/2+2nL\leq y_{12}< L/2+2nL$, and is $-1$ otherwise. The probability distribution for $y_{12}$ is $\int dy_2  \, p(y_{12}-y_2)p(y_2) = e^{-y_{12}^2/8 \gamma(0)t}/\sqrt{8\pi\gamma(0)t}$, which gives
\begin{equation}\label{P_pair}
\langle Z_{\rm ec} \rangle =\sum_{n=-\infty}^{+\infty}(-1)^n {\rm erf}\left( \frac{2n+1}{2\sqrt{2\gamma(0)t/L^2}}\right).
\end{equation}
Both functions (\ref{P0_pair}) and (\ref{P_pair}) are plotted in
Fig.~\ref{pair_of_anyons} and show perfect agreement with the
numerical simulation. An important feature of the above analytical
expressions is that the time dependence only enters through the
combination $\gamma(0)t/L^2$, which makes it possible to scale
curves from different system sizes and diffusion constants onto each
other.

\section{Effective Hamiltonian via Schrieffer-Wolff transformation} \label{app:SW_trafo}

In order to find an effective Hamiltonian for Eq.~\eqref{eq:two-photon coupling},
we write $H = H_0 + V$, where $H_0 = \sum_{i = 1}^2 \omega_i a_i^\dag a_i$ and $V = \sum_p g_p n_p(a_1^\dag a_2 + a_1 a_2^\dag)$, and
treat $V$ as a small perturbation. The general expression for the Schrieffer-Wolff transformation of $H$ up to second order in $V$ reads
\begin{equation}\label{eq:general SW trafo formula}
H_{\rm eff} = H_0 + \frac{i}{2}\lim_{\varepsilon\rightarrow 0}\int_0^\infty dt e^{-\varepsilon t}\left[ V, V(t)\right] + \mathcal{O}(V^3),
\end{equation}
where $V(t) = \exp(iH_0 t)V \exp(-iH_0 t)$, which yields in our case
\begin{equation}
V(t) =  \sum_p g_p n_p\left(e^{i(\omega_1 - \omega_2)t} a_1^\dag a_2 + e^{-i(\omega_1 - \omega_2)t} a_2^\dag a_1\right).
\end{equation}
With this, the commutator in Eq.~\eqref{eq:general SW trafo formula} evaluates to 
\begin{equation}
 [V, V(t)] = 2i(\sum_p g_p n_p)^2 (a_2^\dag a_2 - a_1^\dag a_1)\sin(\omega_1 - \omega_2)t.
\end{equation}
Inserting this into Eq.~\eqref{eq:general SW trafo formula} and performing the integral yields Eq.~\eqref{eq:effective interaction hamiltonian}.





\begin{thebibliography}{38}
\bibitem{footnote01} $\sigma_z$-errors are irrelevant for the dynamics of plaquettes.
\bibitem{footnote02} Note that in Eq. \ref{rate_eq_pairs_int}, we have approximated the number of spins without errors by the total number $2L^2$, neglecting corrections of order $\ln L$.
\bibitem{comment_stars}
{As a simple example, we note that since the gap of the star anyons is also given by Eq.~(\ref{J0_from_Jk}), an additional term $\sum_{s} g_s n_s (a_1^\dag a_2+a_2^\dag a_1)$ appears in Eq.~(\ref{eq:two-photon coupling}) which generates repulsive star-star as well as star-plaquette interactions.}
\end{thebibliography}
\end{document}